%% file: iclr2026_conference.tex
\documentclass{article} 
\usepackage{iclr2026_conference,times}

\input{math_commands.tex}

\usepackage{hyperref}
\usepackage{url}

\usepackage{booktabs}
\usepackage{amsmath}
\usepackage{tabularx}
\usepackage{graphicx}
\usepackage{subcaption}
\usepackage{rotating}
\usepackage{tikz}
\usepackage{tabularx}
\usepackage{makecell}

\title{Persona Alchemy: Designing, Evaluating, and Implementing Psychologically-Grounded LLM Agents for Diverse Stakeholder Representation}

\author{Sola Kim\\
School of Sustainability\\
Arizona State University\\
Tempe, AZ 85281, USA \\
\texttt{\{sola\}@asu.edu} \\
\AND
Dongjune Chang \\
Applied Materials Division \\
Argonne National Laboratory \\
Lemont, IL 60439, USA \\
\texttt{\{changd\}@anl.gov} \\
\AND
Jieshu Wang \\
Department of Technology and Society \\
Stony Brook, NY 11794, USA \\
\texttt{\{jieshu.wang\}@stonybrook.edu}
}

\iclrfinalcopy 
\begin{document}

\maketitle

\begin{abstract}
Despite advances in designing personas for Large Language Models (LLM), challenges remain in aligning them with human cognitive processes and representing diverse stakeholder perspectives. We introduce a Social Cognitive Theory (SCT) agent design framework for designing, evaluating, and implementing psychologically grounded LLMs with consistent behavior. Our framework operationalizes SCT through four personal factors (cognitive, motivational, biological, and affective) for designing, six quantifiable constructs for evaluating, and a graph database-backed architecture for implementing stakeholder personas. Experiments tested agents' responses to contradicting information of varying reliability. In the highly polarized renewable energy transition discourse, we design five diverse agents with distinct ideologies, roles, and stakes to examine stakeholder representation. The evaluation of these agents in contradictory scenarios occurs through comprehensive processes that implement the SCT. Results show consistent response patterns ($R^2$ range: $0.58-0.61$) and systematic temporal development of SCT construct effects. Principal component analysis identifies two dimensions explaining $73\%$ of variance, validating the theoretical structure. Our framework offers improved explainability and reproducibility compared to black-box approaches. This work contributes to ongoing efforts to improve diverse stakeholder representation while maintaining psychological consistency in LLM personas.
\end{abstract}

\clearpage
\section{Introduction}

Natural Language Processing (NLP) technologies have evolved language models into powerful tools, yet their impact on complex societal issues across disciplines remains unclear. Diverse expertise is crucial for evaluating their effectiveness beyond technical benchmarks. Current NLP agents, mostly powered by LLMs, dependent on static prompts, struggle to mimic human-like behavior and longitudinal interaction accurately. Integrating psychological frameworks can enhance NLP systems by modeling human cognition, social dynamics, and decision-making, leading to better diverse stakeholder representation in interdisciplinary environments. We propose that LLM agents grounded in psychological frameworks would provide a novel approach to enhance stakeholder representation in interdisciplinary contexts. Our research explores how these agents are designed and validated, contributing to measuring NLP’s cross-disciplinary impact in three ways: demonstrating how psychological theories inform LLM application design and evaluation; providing evidence that interdisciplinary design principles yield measurable outcomes; and offering an integrated methodology for designing and evaluating LLMs even in transdisciplinary domains like sustainability.

\section{Background and Related Work}

\textbf{Designing} personas, predefined personality profiles guiding dialogue model responses, in LLMs currently presents challenges in aligning with realistic human cognition and personality traits. The presumption of equivalence between language proficiency and thought may overestimate reasoning capabilities \citep{mahowaldDissociatingLanguageThought2024}. Recent NLP systems incorporate psychological theories but only observe effects, not explain causation \citep{sharmaGenerativeEchoChamber2024, phelpsMachinePsychologyCooperation2025}. LLMs can approximate social behaviors but lack psychological plausibility in representing human motivations and their influence on decision-making processes. Traditional personality theories have been applied to understand and measure human personality dimensions \citep{serapio-garciaPersonalityTraitsLarge2025a, hilliardElicitingPersonalityTraits2024}, but they provide limited insight into how behavior is shaped through interactions.

\textbf{Evaluating} persona effects is challenging. Prompting techniques \citep{huQuantifyingPersonaEffect2024} and structured methods like persona codebooks \citep{tangMORPHEUSModelingRole2024, tsengTwoTalesPersona2024} offer frameworks but lack flexibility and generalizability. Researchers struggle to create metrics that balance consistency and adaptability. \citet{haCloChatUnderstandingHow2024} introduced customizable options, but they lacked coherent persona grounding, causing contextually unstable outputs. This becomes problematic due to evolving conversation topics \citep{fischerPersonalityDifferencesDrive2024, templetonConversationalLaunchPads2024}.

\textbf{Implementation} challenges exacerbate theoretical and evaluative shortcomings. Current LLMs use fixed personas that hinder adaptation to evolving user needs, requiring detailed prompt engineering for customization. Persona-conditioning methods are inconsistent and ineffective \citep{giorgiModelingHumanSubjectivity2024}. Even successful implementations raise ethical concerns as simple API-level instructions can significantly alter user perception \citep{deshpandeAnthropomorphizationAIOpportunities2023}. Dataset construction challenges persist, with many systems relying on social media or crowdsourcing, limiting representativeness \citep{leeStarkSocialLongTerm2024, kimSODAMillionscaleDialogue2023}. \citet{bowdenActiveListeningPersonalized2024} developed a large dataset of personalized Q\&A pairs, but it was too large for many research applications. Fine-tuning individual LLMs remains computationally expensive and infeasible at scale. Non-static persona implementation based on in-dialogue among agents has not been fully tested with highly diverse or conflicting personas \citep{chengInDialoguesWeLearn2024}. Key challenges include balancing personalization depth with response diversity \citep{tangMORPHEUSModelingRole2024}, maintaining coherence across sessions \citep{giorgiModelingHumanSubjectivity2024}, and representing perspectives effectively.

\textbf{Representation} issues reveal fundamental limitations in current persona approaches. Persona-conditioning methods inadequately represent underrepresented populations \citep{santurkarWhoseOpinionsLanguage2023}, constraining social science applications. \citet{wangLargeLanguageModels2025} criticized how LLMs misportray marginalized groups by reflecting out-group stereotypes rather than authentic in-group perspectives. Substituting human participants with AI models fundamentally undermines representation, inclusion, and understanding \citep{agnewIllusionArtificialInclusion2024}. Current prompt-based representation methods rely excessively on base models without addressing deeper representation issues \citep{liuEvaluatingLargeLanguage2024, liCAMELCommunicativeAgents2023}. Effective representation requires structural changes to model design and training methodologies rather than superficial prompt engineering.

\section{Design}

\begin{figure*}[!ht]
    \centering
    \includegraphics[width=\textwidth]{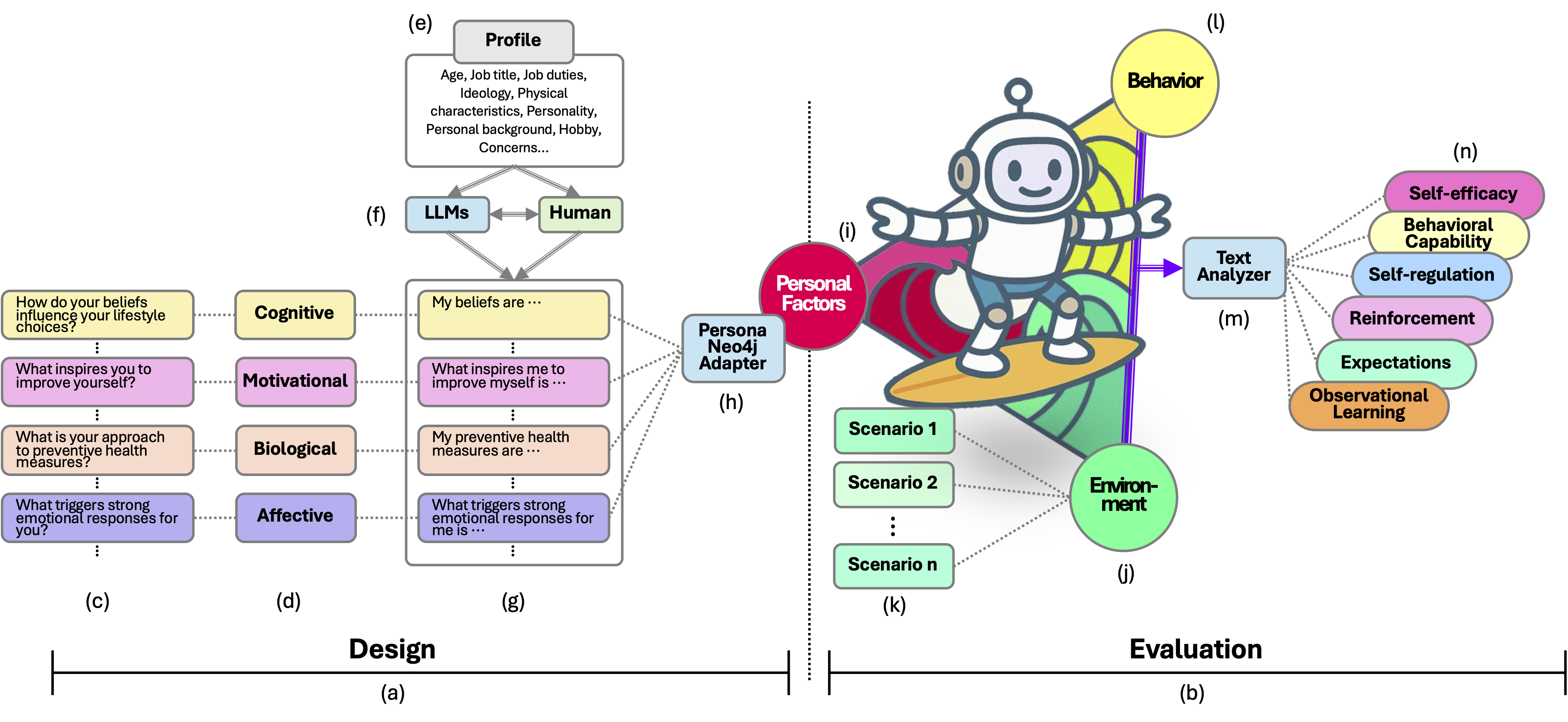}
    \caption{SCT Framework Using Personal Factors for Agent Design and Six Constructs for Cross-Scenario Evaluation within Triadic Reciprocal Determinism. \textit{Note:} Light blue round squares indicate LLMs in the framework.}  
    \label{fig:triadic}
\end{figure*}

\subsection{Social Cognitive Theory Fundamentals}

Social Cognitive Theory \citep{banduraSelfSystemReciprocal1978, banduraSocialFoundationsThought1986, banduraHumanAgencySocial1989,banduraSocialCognitiveTheory2001, banduraSocialCognitiveTheory2023} emphasizes how people learn through observation, experience, and environmental influences. At its core, SCT views humans as active agents who both influence and are influenced by their surroundings, rather than passive recipients of environmental forces. In everyday terms, SCT explains why we might adopt behaviors we see succeed in others, how our beliefs about our capabilities affect our choices, and why the same person might act differently in various social contexts. SCT has broad applications in education \citep{burneyApplicationsSocialCognitive2008, bembenuttyApplyingSocialCognitive2016, schunkSocialCognitiveTheory2001}, organizational behavior \citep{banduraOrganisationalApplicationsSocial1988, ozyilmazTrustOrganizationModerator2018}, mass communication \citep{banduraSocialCognitiveTheory2001a, fuTestingTheoreticalModel2009}, and health \citep{banduraHealthPromotionPerspective2000, godinHealthcareProfessionalsIntentions2008, beauchampSocialCognitiveTheory2019}.

SCT addresses limitations in current LLM persona approaches by moving beyond static prompts and traditional personality theories. Unlike fixed AI personas, SCT creates dynamic agents that evolve through interactions, similar to human development. This framework solves implementation challenges like inconsistent persona-conditioning and adapting to evolving contexts. For example, an SCT-based agent adjusts its reasoning based on new information and social context, rather than simply stating generic role-aligned viewpoints. To design psychologically grounded LLM agent personas, we ground our multi-LLM agent framework in SCT's "triadic reciprocal determinism" \citep{banduraSocialCognitiveTheory2023}. As illustrated in Figure~\ref{fig:triadic}, SCT integrates personal factors (i), environment (j), and behavior (l) to enable psychologically plausible representation of diverse stakeholders. Our agents balance internal beliefs (personal factors) with external information (environment) to produce contextually appropriate responses (behavior), enabling realistic simulation with longitudinal interaction and dynamic adaptation while maintaining psychological coherence.

\subsection{SCT-Based Agent Design Framework Overview}

Our agent design (Figure~\ref{fig:triadic}) combines four personal factors (d) (cognitive, motivational, biological, and affective) with six SCT constructs (n) (self-efficacy, behavioral capability, self-regulation, reinforcement, expectations, and observational learning) to create psychologically grounded agents with diverse stakeholder perspectives and consistent behavior. Scenarios (k) serve as the environment (j), enabling agents to respond contextually while maintaining psychological fidelity. Continuous feedback loops influence behaviors and the environment, generating dynamic interactions that enhance realism in complex social contexts.

\subsection{Personal Factors Operationalization}
The personal factors adapted from SCT (Table \ref{tab:dimensions-questions}) include: cognitive factors (belief structures, knowledge base, and attitudes such as views on individual rights versus common good); biological factors (physical characteristics and demographic information relevant to self-concept); affective factors (emotional tendencies and feeling states influencing information processing and decision-making); and motivational factors (internal drives and goals directing behavior). These personal factors are implemented as a question-and-answer dataset that forms the foundation of each agent's persona. We created 550 balanced questions (Figure \ref{fig:triadic}(c)) covering four categories (Table \ref{tab:dimensions-questions} and Appendix~\ref{app:dataset}) and diverse dimensions like personal identity and social issues. Answers are generated using each agent's profile, developed with diverse perspectives through LLMs. We use a novel-writing framing technique to simulate real-world interview annswers and elicit detailed personas, maintaining consistency across stakeholder types. Our process, illustrated in Figure \ref{fig:triadic}(f-g), includes: prompting LLMs to generate responses by framing each query as "given this character's profile, how would they answer this question"; using multiple language models to address single-model biases; and verification through both an LLM and two human coders, with conflicts resolved by majority rule. This methodology ensures consistent agent personas across diverse interactions.

\begin{table}[!]
\centering
  \resizebox{0.9\textwidth}{!}{%
    \begin{tabular}{lp{5cm}p{8cm}}
      \toprule
    \textbf{Category} & \textbf{Short Definition} & \textbf{Sample Questions} \\
    \midrule
    Cognitive & Mental processes like thinking, reasoning, and understanding that shape perception and learning. & 
    •\;What core values define your personal identity?\newline
    •\;How do your beliefs influence your lifestyle choices?\newline
    •\;In what ways does your self-perception affect your daily decisions? \\
    \midrule    
    Motivational & Internal drives and goals that direct behavior, especially self-efficacy and outcome expectations. &
    •\;What inspires you to improve yourself?\newline
    •\;How do you overcome procrastination in your personal life?\newline
    •\;What drives you to maintain or change your habits? \\
    \midrule    
    Biological & Genetic and physiological factors affecting behavior, considered in interaction with cognition and environment. &
    •\;How has your biological heritage shaped your lifestyle?\newline
    •\;What is your approach to preventive health measures?\newline
    •\;How do you integrate physical health into your personal goals? \\
    \midrule    
    Affective & Emotions and feelings that influence responses, motivation, and decision-making. &
    •\;How does emotional well-being factor into your lifestyle choices?\newline
    •\;What role does emotional intelligence play in your personal growth?\newline
    •\;What triggers strong emotional responses for you?\\    
      \bottomrule
    \end{tabular}
  }
  \caption{Categories of Social Cognitive Theory and Sample Questions for Constructing Datasets}\label{tab:dimensions-questions}
\end{table}

\subsection{Implementation}
We used Neo4j-backed graph database system \citep{neo4jincNeo4jGraphsEveryone2025} to store personal factors for each agent persona. Each agent is powered by \texttt{Llama-3.2-3B-Instruct} as the base language model \citep{metaaiLlama32Revolutionizing2024}. The system organizes persona data hierarchically through \texttt{Agent-Category-Dimension-Question} relationships, allowing contextual retrieval of relevant information during conversations. The \texttt{PersonaNeo4jAdapter} (Figure~\ref{fig:triadic}(h)) imports personal factors from JSON datasets and retrieves agent-specific information via Cypher queries, using the \texttt{mxbai-embed-large-v1} embedding model for semantic similarity \citep{leeOpenSourceStrikes2024}. During message processing, the architecture extracts relevant categories from incoming messages and retrieves corresponding personal factors to compile a background section for the language model prompt, ensuring relevant responses by incorporating only personal information relevant to the conversation topic.

\section{Evaluation}

\subsection{SCT Constructs as Evaluation Metrics}

We implement SCT's six core constructs as quantifiable metrics to assess (Figure~\ref{fig:triadic}(n)) how consistently agent personas respond when faced with contradicting information, regardless of the specific scenario context (Figure~\ref{fig:triadic}(j-k)). Each SCT construct serves as a distinct dimension for evaluating persona consistency (Table~\ref{tab:eval-criteria}).

\begin{table}[!htb]
\centering
\resizebox{0.9\textwidth}{!}{%
\begin{tabular}{p{2.5cm}p{4.5cm}p{8cm}}
\toprule
\textbf{Construct} & \textbf{Definition} & \textbf{Evaluation Criteria} \\
\midrule
Self-Efficacy & Agent's confidence in its abilities to achieve goals and influence outcomes. &
•\;Maintains appropriate confidence levels when challenged\newline
•\;Demonstrates resilience in response to contradicting information \\
\midrule
Behavioral Capability & Knowledge and skills necessary to perform specific behaviors. &
•\;Exhibits domain-appropriate expertise when challenged\newline
•\;Applies skills consistently across contradicting scenarios \\
\midrule
Expectations & Anticipated outcomes of behaviors and consequence evaluations. &
•\;Maintains consistent outcome predictions\newline
•\;Shows stable risk-benefit assessments when faced with contradictions \\
\midrule
Self-Regulation & Ability to monitor and maintain goal-directed behaviors despite challenges. &
•\;Demonstrates consistent self-monitoring processes\newline
•\;Shows appropriate adjustment behaviors while preserving core values \\
\midrule
Observational Learning & Process of incorporating information gained from observing others. &
•\;Shows consistent patterns in information selection\newline
•\;Maintains perspective consistency when learning from contradicting sources\\
\midrule
Reinforcements & Responses to positive or negative feedback, rewards, or consequences that influence future behavior. &
•\;Demonstrates increased motivation following positive feedback or rewards\newline
•\;Expresses fulfillment or obligation as a result of reinforcement experiences \\
\bottomrule
\end{tabular}
}
\caption{Social Cognitive Theory Constructs and Evaluation Criteria for Agent Persona Assessment}\label{tab:eval-criteria}
\end{table}

\subsection{Evaluation Operationalization}

Our methodology provides a domain-independent approach to persona evaluation through a systematic five-step process (Figure~\ref{fig:triadic}(e-n)). First, we establish initial SCT construct profiles for each persona. Second, we develop contradictory scenarios. Third, we analyze responses through all six SCT construct dimensions to measure consistency. Fourth, we compare responses against expected persona-consistent patterns. Finally, we track SCT construct expression changes across interaction rounds to evaluate temporal development. This framework supports quantitative assessment of persona consistency across diverse contexts. We quantify SCT construct expression on continuous scales (0.1 to 1.0), with higher values indicating better alignment with exemplars. The evaluation references comprehensive configuration examples illustrating varying levels of each SCT construct (detailed in Appendix~\ref{app:eval-config}).

\subsection{Implementation}

We implemented our agent evaluation using Neo4j, encoding six SCT constructs as quantifiable parameters within each agent persona (Table~\ref{tab:eval-criteria}). The \texttt{TextAnalyzer} component (Figure~\ref{fig:triadic} (m)) uses \texttt{Llama-3.2-3B-Instruct} \citep{metaaiLlama32Revolutionizing2024} to analyze information across semantic, emotional, and SCT construct alignments. The system integrates Retrieval-Augmented Generation (RAG) with \texttt{PersonaNeo4jAdapter} to access persona information and enable persona-consistent responses. Our framework supports various evaluator types (LLMs, human experts, specialized algorithms) and involves recording responses during contradictory scenarios, analyzing them against construct exemplars, assigning normalized scores, and tracking temporal development through repeated evaluations.

\section{Simulated Case study: Renewable Energy Transition Discourse among Diverse Stakeholders}

\subsection{Background}

Our research uses renewable energy transition discourse as a test case for diverse stakeholder representation due to its cultural and group identity-based polarization \citep{kahanGeoengineeringClimateChange2015, hartBoomerangEffectsScience2012}. In energy transition discussions, stakeholders interact in complex negotiations with conflicting information. Renewable energy is a stakeholder issue \citep{ruggieroRealizingSocialAcceptance2014} with persistent conflicts over its socio-political space \citep{lauberPoliticsEconomicsConstructing2016}. Stakeholders with diverse ideological stances must navigate conflicting claims about economic impacts, environmental consequences, and technological feasibility. Our SCT-based evaluation framework assesses how consistently these diverse stakeholders maintain their positions with conflicting information.

\subsection{Personal Factors: Agent's Persona} \label{simul:persona}

We developed five diverse agent profiles (See Figure \ref{fig:agent-profile} for the details) with varying ideological orientations using GPT-4 \citep{openaiGPT4TechnicalReport2024} via ChatGPT \citep{openaiIntroducingChatGPT2022}, representing diverse stakeholders in energy transition discussions. Using ChatGPT, we controlled only the ideology of agents by creating novel's characters with different stakes in the renewable energy transition, allowing other personality aspects to emerge naturally. Each profile included comprehensive attributes: name, age, job title, ideology, physical characteristics, personality, personal background, job duties, hobbies, and concerns. We prompted LLMs to generate profile-consistent responses to 550 pre-defined questions using multiple language models: GPT-4o-mini \citep{openaiHelloGPT4o2024}, Mistral-7B-v0.1 \citep{jiangMistral7B2023a}, and zephyr-7b-alpha \citep{tunstallZephyrDirectDistillation2023}. For each question, we used the framing "Given this character X's profile, how would they answer this question?" to elicit authentic, persona-consistent responses. The responses were verified by the authors and GPT-4o to ensure consistency and accuracy. Human verification involved qualitative assessment of persona alignment, ensuring responses authentically reflected stakeholder perspectives and internal consistency with character profiles. 


\subsection{Environment: Contradicting Information Scenarios}
To assess agent persona consistency in triadic reciprocal determinism (Figure~\ref{fig:triadic}), we designed contradictory information scenarios (k) challenging each agent’s core personal factors (i) and behaviors (l). These scenarios included foundational beliefs, counter-evidence, varying reliability, and domain relevance. For instance, Douglas Harrington (coal mining CEO) faced statements about renewable energy job creation and coal’s economic disadvantages, while Sierra Jameson (renewable energy consultant) encountered contradictory information about solar panel carbon footprints and reliability issues. Each contradictory statement contained hidden reliability metadata, with higher values representing well-supported information. This reliability range matched real-world information evaluation patterns: highest values for peer-reviewed scientific studies, moderate values for government reports, and lower values for non-peer-reviewed sources. This variation tested whether agents calibrated responses based on information quality. Each agent encountered five distinct scenarios presented sequentially with increasing complexity across multiple interaction rounds. After each presentation, we recorded responses for subsequent six SCT construct evaluation, revealing how different constructs manifested when agents navigated information challenging their personas.

\subsection{Experimental Setup}
Our experiment involved 5 interaction rounds where agents faced contradictory facts challenging their mental representations, with 100 iterations per condition for statistical validity. We presented scenarios with factual assertions that either aligned or contradicted the agent’s beliefs. Each scenario included domain-specific information and strategically positioned contradictory elements to challenge the agent’s core beliefs. The contradictions were calibrated to maintain plausibility and trigger belief reconciliation processes. Comprehensive analyses (bootstrap confidence intervals, round subset sensitivity, leave-one-out testing; Appendix ~\ref{app:sensitivity}) confirmed our design's validity and showed consistent effect size progression across rounds. Response patterns were measured through automated content analysis of agent outputs, tracking changes in certainty markers, reference to prior beliefs, incorporation of new information, and justification strategies—providing quantitative metrics of cognitive adaptation processes. The agent architecture leverages a neuroscience-inspired Enhanced Memory System \citep{changOutStatesMind2025} with multi-type memory, RAG-based retrieval, dynamic SCT constructs tracking, and source reliability integration—modeling realistic cognitive processes (Appendices ~\ref{app:llm}, ~\ref{app:eval-config}).

\section{Results}

\subsection{Model Specification}

We modeled SCT-based response patterns as a function of contradictory information scenarios and SCT constructs using two hierarchical linear models: a fixed-effects model (Model 1) and a time-varying model (Model 2). Both models were estimated with random intercepts for each iteration. A likelihood ratio test compared the models, yielding $\Lambda = 399.82$ ($p < .001$), suggesting temporal interactions with SCT constructs improve model fit. Details are in Appendix~\ref{app:model-comparision}.







\subsection{Behavior: Response Patterns to Contradicting Information Scenarios}

Our analysis reveals consistent agent responses to contradictory information aligned with model specifications. Table~\ref{tab:contradiction-reliability} shows Model~1 (fixed-effects) results where contradicting information consistently predicted SCT-based response patterns. The coefficient ($\beta_1$) remained stable ($1.71$--$1.74$) with high explanatory power ($R^2$: $0.58$--$0.61$).

SCT-based agents demonstrated substantially stronger responses to contradicting information compared to the vanilla agent (coefficient $\sim 1.73$ vs.\ $0.36$), a nearly 5-fold increase. The vanilla agent's higher $R^2$ ($0.83$) coupled with its lower coefficient suggests more rigid, less psychologically plausible belief dynamics than our SCT-based implementation.

\begin{table}[!htbp]
\centering
  \begin{tabular}{lccc}
    \toprule
    \textbf{Agent} & \textbf{Coefficient (SE)} & \textbf{$R^2$} & \textbf{95\% CI} \\
    \midrule
    Douglas Harrington & 1.72 (0.065)$^{***}$ & 0.58 & [1.59, 1.85] \\
    Elizabeth Montgomery & 1.73 (0.063)$^{***}$ & 0.60 & [1.61, 1.85] \\
    Michael Donovan & 1.71 (0.064)$^{***}$ & 0.59 & [1.58, 1.83] \\
    Sierra Jameson & 1.74 (0.065)$^{***}$ & 0.59 & [1.61, 1.87] \\
    Tayen Kaya & 1.73 (0.061)$^{***}$ & 0.61 & [1.61, 1.85] \\
    Vanilla Agent & 0.36 (0.008)$^{***}$ & 0.83 & [0.34, 0.37] \\    
    \bottomrule
  \end{tabular}
\\
\vspace{3pt}
\footnotesize \textit{Note:} $^{***}p < 0.001$
\caption{Fixed-Effects Model (Model 1): Contradicting Information Effects by Agent}
\label{tab:contradiction-reliability}
\end{table}

The consistency across agents with different backgrounds confirms our SCT framework successfully implements plausible persona dynamics regardless of stakeholder viewpoint. The mixed-effects version of Model~1 confirmed statistically insignificant agent differences when controlling for contradictory information (all $p > .85$, $\eta^2 = 0.0002$), supporting the $\beta_1$ coefficient stability and framework robustness across persona implementations.

\subsection{Temporal Development of SCT Construct Effects}

Building on the Model 1's results, we examined Model 2 to investigate how SCT constructs' influence changes over time. Table~\ref{tab:temporal-development} presents the estimated parameters from our tempoeral development  model, showing statistically significant interactions ($p < .05$) between SCT constructs and interaction rounds. This confirms our hypothesis that the influence of SCT constructs develops systematically over successive interactions.

\begin{table}[!htbp]
\centering
  \begin{tabular}{lccc}
    \toprule
    Parameter & Coefficient & SE & p-value \\
    \midrule
    Intercept & -0.127 & 0.038 & 0.999 \\
    Source reliability & 1.426 & 0.020 & $<$0.001$^{***}$ \\
    \addlinespace
    \textit{SCT Constructs} \\
    \quad Self-efficacy & 2.510 & 0.220 & 0.999 \\
    \quad Behavioral capability & -0.124 & 0.151 & 0.999 \\    
    \quad Expectations & -1.569 & 0.138 & 0.999 \\  
    \quad Self-regulation & -1.373 & 0.114 & 0.999 \\        
    \quad Observational learning & 1.397 & 0.110 & 0.999 \\
    \quad Reinforcements & -1.871 & 0.151 & 0.999 \\    
    \addlinespace
    \textit{Temporal Development} \\
    \quad Self-efficacy × round & 0.318 & 0.103 & 0.002$^{**}$ \\
    \quad Behavioral capability × round & -0.036 & 0.042 & 0.387 \\
    \quad Expectations × round & -0.211 & 0.059 & $<$0.001$^{***}$ \\    
    \quad Self-regulation × round & -0.135 & 0.050 & 0.007$^{**}$ \\    
    \quad Observational learning × round & 0.115 & 0.055 & 0.035$^{*}$ \\
    \quad Reinforcements × round & -0.172 & 0.077 & 0.025$^{*}$ \\    
    \addlinespace
    Random effects variance & $7.36 \times 10^{-6}$ & -- & 0.951 \\
    Residual variance & 0.035 & -- & -- \\
    \bottomrule
  \end{tabular}
    \\
\vspace{3pt}
\footnotesize \textit{Note}: $^{*}p<0.05$, $^{**}p<0.01$, $^{***}p<0.001$  
    \caption{Temporal Development Effects Model (Model 2) Parameter Estimates}
  \label{tab:temporal-development}
\end{table}

\begin{figure}[!htbp]
  \centering
    \begin{subfigure}[t]{0.49\linewidth}
    \includegraphics[width=\linewidth]{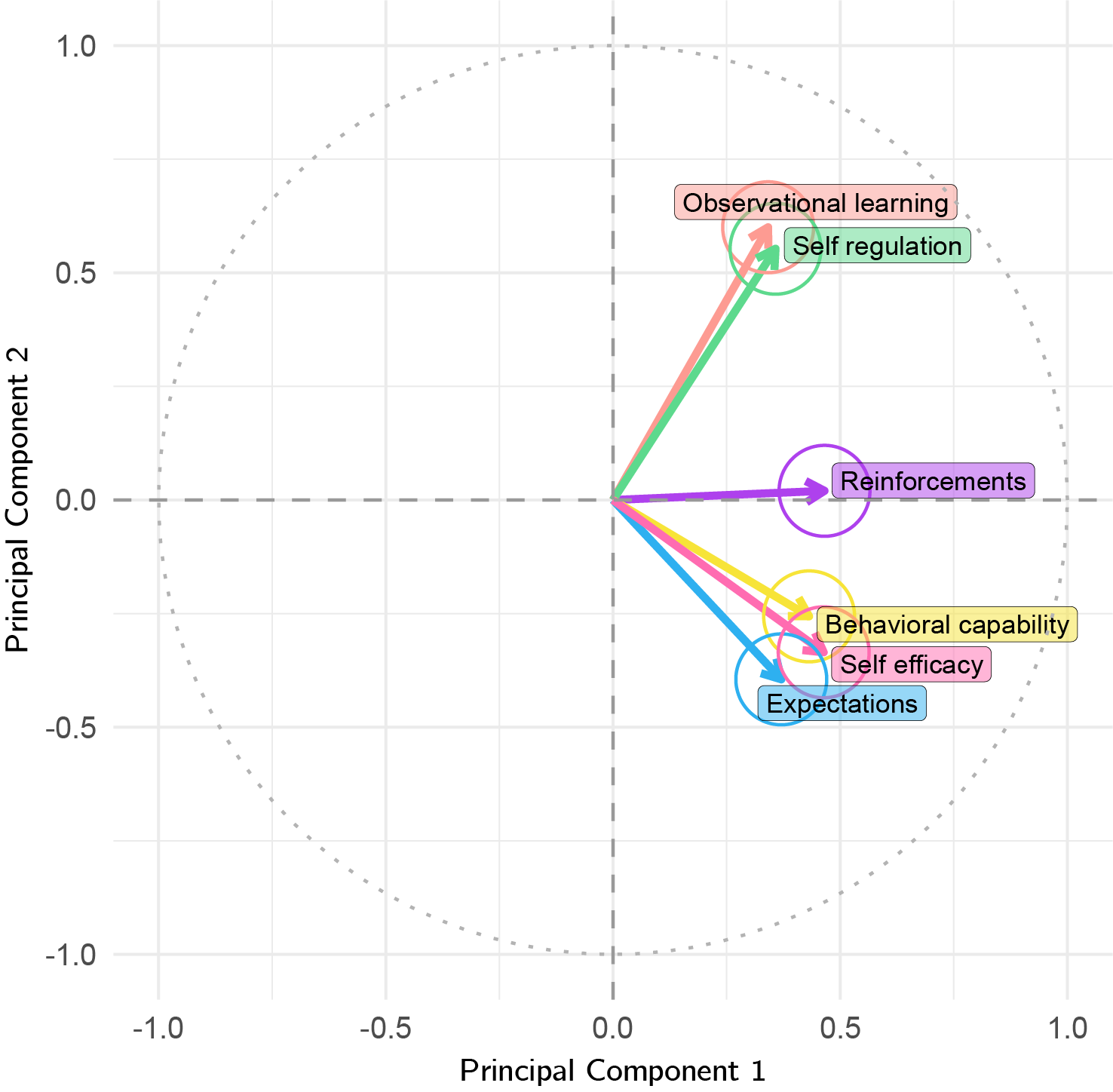}    
    \caption{}
    \label{fig:biplot}    
    \end{subfigure}  
    \hspace{3pt}
    \begin{subfigure}[t]{0.49\linewidth}
    \includegraphics[width=\linewidth]{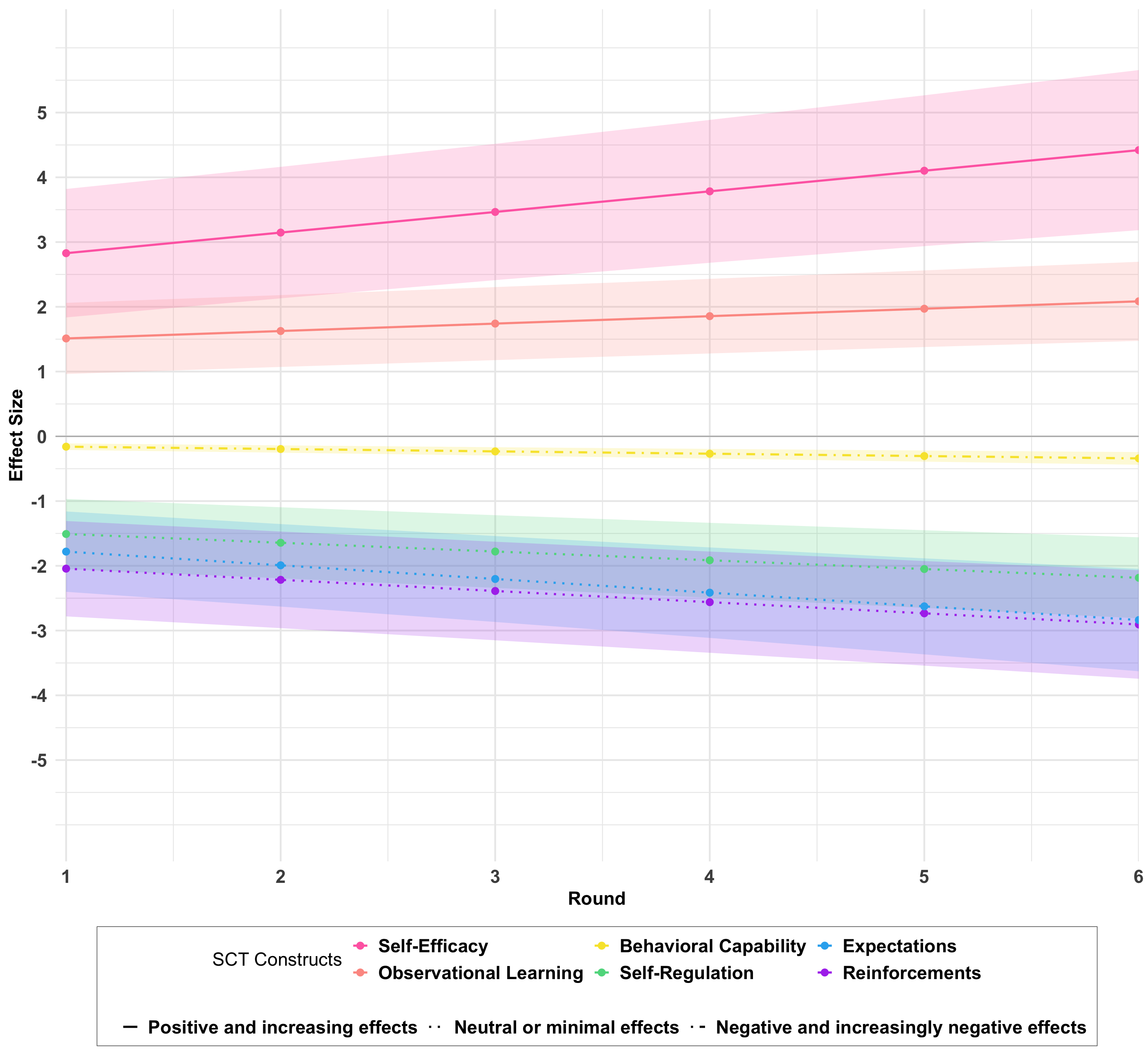}
    \caption{}
    \label{fig:agent-effect}    
    \end{subfigure}   
\caption{Principal component structure of SCT constructs (left) and their temporal effect trajectories across interaction rounds with 95\% confidence intervals (right).}  
\label{fig:overall}    
\end{figure}

Figure~\ref{fig:agent-effect} visualizes SCT construct effects across interaction rounds. Self-efficacy shows the strongest positive trajectory ($\beta = 0.318$, $p = .002$), with agents resisting contradicting information over time. Observational learning follows a positive trajectory ($\beta = 0.115$, $p = .035$), suggesting improved information evaluation with repeated exposure. Conversely, expectations ($\beta = -0.211$, $p < .001$), reinforcements ($\beta = -0.172$, $p = .025$), and self-regulation ($\beta = -0.135$, $p = .007$) negatively affect agents’ susceptibility to response modifications from contradictory information. Behavioral capability remains stable ($\beta = -0.036$, $p = .387$), indicating consistent knowledge application. Positive values represent SCT constructs enhancing resistance to contradicting information, while negative values denote increasing responsiveness. Self-regulation develops temporally, suggesting highly self-regulated agents become more responsive to contradictory information scenarios, which is valuable for evidence-based response adaptation.

Table~\ref{tab:evolution-effects} quantifies the temporal development of each SCT construct across interaction rounds. The magnitude of these changes reveals substantial development, with Self-efficacy showing the strongest positive trajectory ($+1.59$ from Round~1 to Round~6) and Expectations demonstrating the most pronounced negative development ($-1.06$). These quantified changes illustrate how agent response patterns systematically evolve over repeated exposure to contradicting information.


\begin{table*}[!htbp]
\centering
\resizebox{\textwidth}{!}{%
\begin{tabular}{
>{\raggedright\arraybackslash}p{1.8cm}   
>{\raggedleft\arraybackslash}p{2.4cm}    
>{\raggedleft\arraybackslash}p{2.6cm}
>{\raggedleft\arraybackslash}p{2.6cm}
>{\raggedleft\arraybackslash}p{2.3cm}
>{\raggedleft\arraybackslash}p{2.2cm}
>{\raggedleft\arraybackslash}p{2.4cm}
}
\toprule
 & \textbf{Self-Efficacy} & \textbf{Observational Learning} & \textbf{Behavioral Capability} & \textbf{Self-Regulation} & \textbf{Expectations} & \textbf{Reinforcements} \\
\midrule
\multicolumn{7}{l}{\textit{Summary Statistics}} \\
Mean Effect       & 3.62 & 1.80 & $-$0.25 & $-$1.85 & $-$2.31 & $-$2.47 \\
Median            & 3.62 & 1.80 & $-$0.25 & $-$1.85 & $-$2.31 & $-$2.47 \\
Std Dev           & 0.60 & 0.21 & 0.07    & 0.25    & 0.40    & 0.32 \\
Std Error         & 0.56 & 0.29 & 0.04    & 0.29    & 0.35    & 0.40 \\
95\% CI (Lower)   & 2.53 & 1.22 & $-$0.32 & $-$2.42 & $-$3.00 & $-$3.25 \\
95\% CI (Upper)   & 4.72 & 2.37 & $-$0.18 & $-$1.27 & $-$1.62 & $-$1.70 \\
\midrule
\multicolumn{7}{l}{\textit{Temporal Values}} \\
Round 1           & 2.83 & 1.51 & $-$0.16 & $-$1.51 & $-$1.78 & $-$2.04 \\
Round 3           & 3.46 & 1.74 & $-$0.23 & $-$1.78 & $-$2.20 & $-$2.39 \\
Round 6           & 4.42 & 2.09 & $-$0.34 & $-$2.18 & $-$2.84 & $-$2.91 \\
$\Delta$ (R6--R1) & 1.59 & 0.57 & $-$0.18 & $-$0.68 & $-$1.06 & $-$0.86 \\
\bottomrule
\end{tabular}%
}
\\
\vspace{3pt}
\footnotesize \textit{Note:} Standard errors estimated as 20\% of effect size. Results based on 100 iteration experiment.
\caption{Summary Statistics of Temporal Development}
\label{tab:evolution-effects}
\end{table*}

\subsection{Principal Component Analysis (PCA) of SCT Constructs}

PCA \citep{woldPrincipalComponentAnalysis1987} revealed two key components explaining $73\%$ of SCT construct variance. PC1 (eigenvalue=$2.76$, $46\%$ variance) showed positive loadings across all constructs, particularly Self-efficacy $(0.464)$ and Reinforcements $(0.466)$, representing a general "response tendency." PC2 (eigenvalue=$1.62$, $27\%$ variance) differentiated learning-oriented constructs (Observational Learning: $0.600$, Self-regulation: $0.553$) from expectation-based constructs (Expectations: $-0.395$, Self-efficacy: $-0.335$). This component structure aligns with theoretical expectations that cognitive and behavioral aspects of SCT function as distinct but complementary dimensions in agent reasoning.

\begin{table}[!htbp]
\centering
\begin{tabular}{lccc}
\toprule
\textbf{SCT Construct} & \textbf{PC1} & \textbf{PC2} & \textbf{Communality} \\
\midrule
Self-efficacy & \textbf{0.464 }& -0.335 & 0.327 \\
Reinforcements & \textbf{0.466} & 0.020 & 0.217 \\
Behavioral capability & \textbf{0.432} & -0.256 & 0.252 \\
Expectations & 0.370 & -0.395 & 0.293 \\
Self-regulation & 0.358 & \textbf{0.553} & 0.434 \\
Observational learning & 0.342 & \textbf{0.600} & 0.477 \\
\midrule
Eigenvalue & 2.76 & 1.62 & \\
Variance explained & 46\% & 27\% & \\
Cumulative variance & 46\% & 73\% & \\
\bottomrule
\end{tabular}
\\
\vspace{3pt}
\raggedright \footnotesize 
\textit{Note:} Principal Component Analysis with Varimax and Kaiser Normalization. Significant loadings ($\geq0.40$) are shown in bold.
\caption{Principal Component Analysis of SCT Constructs} \label{tab:pca-results}
\end{table}
The vector relationships visible in Figure~\ref{fig:biplot} further illuminate how constructs operate together. Closely aligned vectors like Self-regulation and Observational Learning indicate these constructs frequently co-occur in agent responses, while the near-orthogonal relationship between Reinforcements and Observational Learning suggests these constructs operate relatively independently. This empirical structure provides valuable insight into how cognitive mechanisms interact when agents evaluate contradicting information of varying reliability.

\section{Conclusion}

Our SCT-based framework for LLM agent personas significantly advances current approaches by providing psychologically grounded, consistent stakeholder representations that balance stability and adaptability. We demonstrated how this approach maintains persona consistency while allowing responsiveness to varying information reliability.

Our findings reveal distinct patterns in SCT construct evolution, with self-efficacy showing positive growth and reinforcements showing negative effects. These insights inform nuanced agent design across contexts. Psychological grounding is crucial for authentic stakeholder representation, and temporal development patterns reveal systematic persona evolution while preserving fidelity. Our methodology provides a reproducible framework for evaluating persona consistency beyond renewable energy.

As LLMs mediate human communication and decision-making, psychologically plausible persona frameworks will be critical for responsible technology development. Future work integrating this approach with complementary psychological models and multi-agent interactions will enhance artificial social systems in education, healthcare, and sustainable development.

\subsubsection*{Acknowledgments}
We appreciate the constructive feedback from Gil Speyer, Hyun Lee, Joy Ming, and Yi Ding, as well as the anonymous reviewers. The authors acknowledge Research Computing at Arizona State University \citep{jenneweinSolSupercomputerArizona2023} for providing HPC resources that have contributed to the research results reported within this paper.

\bibliography{bib}
\bibliographystyle{iclr2026_conference}

\appendix

\begin{sidewaysfigure}
    \includegraphics[width=\textwidth]{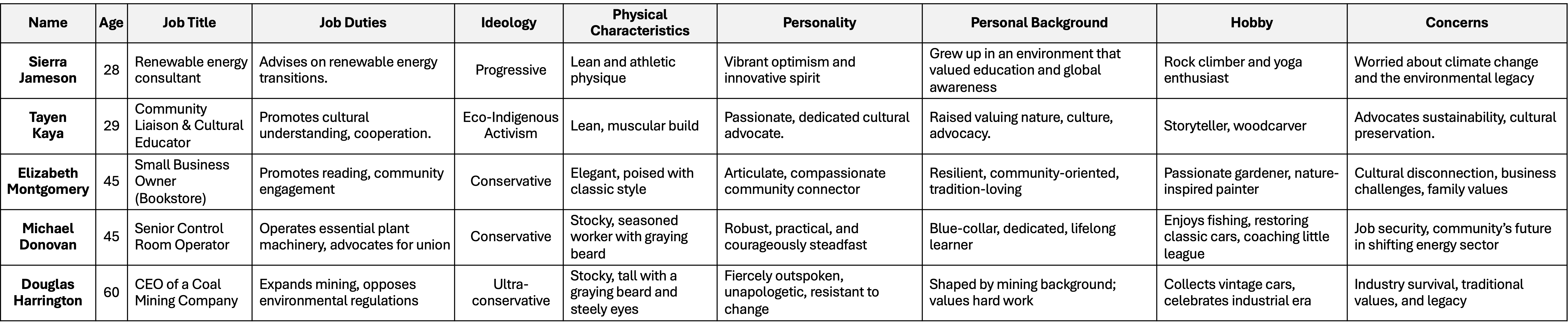}
    \caption{SCT Agent's Profiles for Personas}
    \label{fig:agent-profile}
\end{sidewaysfigure}

\section{Dataset}\label{app:dataset}
\input{appendix/dataset}

\section{LLM Configuration} \label{app:llm}
\input{appendix/llm}

\section{Evaluation Configuration}\label{app:eval-config}
\input{appendix/eval_configuration}

\section{Model Equations}\label{app:models}
\input{appendix/models}

\section{Robustness and Sensitivity Analyses} \label{app:sensitivity}
\input{appendix/sensitivity_analaysis}

\end{document}

%% file: math_commands.tex
\usepackage{amsmath,amsfonts,bm}

\def\eqref#1{equation~\ref{#1}}

\def\1{\bm{1}}

\DeclareMathAlphabet{\mathsfit}{\encodingdefault}{\sfdefault}{m}{sl}
\SetMathAlphabet{\mathsfit}{bold}{\encodingdefault}{\sfdefault}{bx}{n}



%% file: appendix/dataset.tex
For peer review purposes, our agent persona configurations and dataset structure are included in the review submission as a separate file.

%% file: appendix/llm.tex
To ensure reproducibility and transparency, we detail the full language model configuration used in our agent-based experimental framework. The configuration was designed to prioritize character consistency and psychological plausibility.

\subsection{Model and Implementation}
\begin{itemize}
\item \textbf{Primary Model:} \texttt{Llama-3.2-3B-Instruct} \cite{metaaiLlama32Revolutionizing2024} was chosen for several reasons: (1) strong instruction-following for consistent agent personas, (2) optimal balance between performance and efficiency for running multiple instances, (3) superior persona consistency compared to larger models with excessive creativity, and (4) deterministic outputs when temperature is constrained, crucial for experimental reproducibility in this research.

\item \textbf{Framework:} Langchain \cite{chaseLangChain2022} and Ollama \cite{ollamaOllama2025}
\item \textbf{LLM Instances:} Separate LLMs are used for persona interaction and analytical evaluation.
\end{itemize}

\subsection{Hyperparameters}
\begin{itemize}
\item \textbf{Temperature:} Set to $0.2$ to ensure consistent, deterministic outputs aligned with stable agent behavior.
\item \textbf{Token Management:} Default Ollama settings; managed automatically by the Langchain/Ollama integration.
\item \textbf{Sampling:} Top-k and top-p sampling were not used. This is justified because the low temperature $(0.2)$ yields a sharply peaked probability distribution, effectively limiting token selection to high-probability outputs. This makes additional sampling constraints unnecessary for maintaining consistent agent personas throughout the experiment.
\end{itemize}

\subsection{Prompt Engineering}
We employ structured \texttt{ChatPromptTemplate} prompts with strict role definitions to enforce character fidelity:

\begin{itemize}
\item \textbf{Persona Initialization Template:} Establishes the agent's identity, background, profession, and values. Prompts emphasize authentic, first-person responses and prohibit breaking character or acknowledging AI identity.
\item \textbf{Conversation Response Template:} Integrates context, persona background, interlocutor identity, and value-oriented instructions to maintain role coherence during interaction.
\end{itemize}

\subsection{Embeddings and Retrieval}
\begin{itemize}
\item \textbf{Embedding Model:} \texttt{mxbai-embed-large} (1024 dimensions) \cite{leeOpenSourceStrikes2024}
\item \textbf{Similarity Metric:} Cosine similarity with a 0.6 threshold
\item \textbf{Purpose:} Memory retrieval and relevance filtering for contextual awareness.
\end{itemize}

%% file: appendix/eval_configuration.tex
This appendix describes the evaluation configuration used in the current study. The research leverages an established memory framework \cite{changOutStatesMind2025} that has been adapted for the purposes of this experiment. While this framework is not the focus of the current study, these configuration details are provided for reproducibility.

\subsection{Social Cognitive Theory Constructs Evaluation}
Agent responses were evaluated across six core SCT constructs using a three-level scoring system (0.0: negative alignment, 0.5: neutral alignment, 1.0: positive alignment):
\begin{itemize}
\item \textbf{Self-Efficacy}: Confidence in one's ability to make a positive environmental impact
\begin{itemize}
\item 0.0: Expresses doubt about personal influence (e.g., "I don't think my individual actions can make a significant difference")
\item 0.5: Shows uncertainty or mixed confidence (e.g., "I might try some eco-friendly habits, but I'm not sure if I'll stick to them")
\item 1.0: Demonstrates strong belief in personal capability (e.g., "I am confident that I can make eco-friendly choices daily")
\end{itemize}
\item \textbf{Behavioral Capability}: Knowledge and skills in sustainable practices
\begin{itemize}
\item 0.0: Shows lack of knowledge or skills (e.g., "I don't really know how to start saving energy")
\item 0.5: Demonstrates partial knowledge or uncertainty (e.g., "I know a bit about saving energy but not always sure how to implement")
\item 1.0: Exhibits mastery of sustainable practices (e.g., "I've learned how to save energy effectively")
\end{itemize}
\item \textbf{Expectations}: Beliefs about the impact of one's actions
\begin{itemize}
\item 0.0: Shows pessimism about impact (e.g., "I don't think my actions will make a difference")
\item 0.5: Expresses mixed or uncertain expectations (e.g., "It could be good to try, but I'm not expecting big impact")
\item 1.0: Maintains positive expectations about outcomes (e.g., "I expect to significantly reduce my environmental impact")
\end{itemize}
\item \textbf{Self-Regulation}: Ability to set and maintain environmental goals
\begin{itemize}
\item 0.0: Exhibits poor self-monitoring or goal-setting (e.g., "I always forget to keep track, so I stopped trying")
\item 0.5: Shows inconsistent self-regulation (e.g., "I sometimes think about monitoring but don't set strict goals")
\item 1.0: Demonstrates strong self-monitoring and goal achievement (e.g., "I regularly check energy usage and set goals monthly")
\end{itemize}
\item \textbf{Observational Learning}: Learning from others' environmental behaviors
\begin{itemize}
\item 0.0: Shows resistance to social modeling (e.g., "Most people I know don't bother, so why should I")
\item 0.5: Demonstrates partial influence from others (e.g., "I see others saving energy and consider maybe I could too")
\item 1.0: Exhibits strong positive influence from others (e.g., "Friends saving energy waste has inspired me to adopt similar habits")
\end{itemize}
\item \textbf{Reinforcements}: Motivation from feedback and outcomes
\begin{itemize}
\item 0.0: Shows lack of motivation from feedback (e.g., "No one noticed or cared, so I didn't continue")
\item 0.5: Exhibits mixed response to reinforcement (e.g., "I've been praised occasionally, but it hasn't changed habits much")
\item 1.0: Demonstrates strong positive reinforcement effect (e.g., "Positive feedback motivates me to continue and improve")
\end{itemize}
\end{itemize}
The evaluation process involved comparing agent responses against example statements, with an LLM evaluating alignment between responses and category examples to assign scores. Values were tracked over time to measure belief evolution when agents were exposed to contradicting information.

\subsection{Memory Evaluation Scales} \label{app:memory}

All evaluations use 1-7 Likert scales with the following dimensions:

\subsubsection{Short-Term Memory} \begin{itemize} \item \textbf{Agreement}: From 1 (strong disagreement) to 7 (strong agreement) \item \textbf{Impression}: From 1 (very negative: confrontational/dismissive) to 7 (very positive: constructive/empathetic) \item \textbf{Relevance}: From 1 (not relevant) to 7 (highly relevant) \end{itemize}

\subsubsection{Long-Term Memory} \begin{itemize} \item \textbf{Importance}: From 1 (not important) to 7 (critical), threshold: 0.7 for transfer \item \textbf{Persistence}: From 1 (very short-lived) to 7 (permanent) \end{itemize}

\subsubsection{Shared Memory} \begin{itemize} \item \textbf{Consensus}: From 1 (no consensus) to 7 (complete consensus) \item \textbf{Impact}: From 1 (no impact) to 7 (critical impact) \item \textbf{Collaboration}: From 1 (no collaboration) to 7 (full integration) \end{itemize}

\subsection{Weighting Systems} Importance scores are calculated using multiple weighted factors:

\subsubsection{Importance Score Components} \begin{itemize} \item Type Score: 0.4 \item RAG Metrics: 0.3 \item Message Type: 0.1 \item Recency: 0.1 \item Agent Relationship: 0.1 \end{itemize}

\subsubsection{Context-Specific Weights} \begin{itemize} \item \textbf{Message Type}: Spoken (0.7), Heard (0.3), Default (0.5) \item \textbf{Recency}: Current (1.0), Recent (0.8), Older (0.6), Oldest (0.4) \item \textbf{Agent Relationship}: Own (0.7), Other (0.3) \end{itemize}

\subsection{Memory Transfer Rules} \begin{itemize} \item \textbf{Short-Term to Long-Term}: Agreement $\geq$ 6, Impression $\geq$ 6, Relevance $\geq$ 5 \item \textbf{Long-Term to Shared}: Importance $\geq$ 6, Persistence $\geq$ 5, Consensus $\geq$ 6 \end{itemize}

\subsection{RAG System Metrics} The Retrieval-Augmented Generation system evaluates memory using: \begin{itemize} \item \textbf{Memory Trace} (weight: 0.4): Quantifies encoding strength using Alpha (0.6) and Beta (0.4) parameters \item \textbf{Similarity} (weight: 0.4): Measures relationship to existing memories with Lambda factor (0.2) \item \textbf{Interference} (weight: 0.2): Accounts for competing memories with Mu factor (0.3) \end{itemize}

%% file: appendix/models.tex
\subsection{Model Equations} 

\subsubsection{Model 1: Fixed Effects Model}\label{app:model1} 
The fixed-effects model is specified as:

\begin{align}
y_{ijt} =\ & \beta_0 + \beta_1\, \text{C}_{ijt} + \sum_{k=2}^{7} \beta_k\, \text{X}_{ki} + u_j + \varepsilon_{ijt}
\end{align}

\noindent where the variables are defined as follows:
\[
\begin{aligned}
y_{ijt} & : \text{ SCT-based response patterns for agent } i, \\
        & \quad \text{iteration } j, \text{ at round } t \\
\text{C}_{ijt} & : \text{ contradicting information scenarios} \\
\text{X}_{2i} & = \text{Reinforcements}_i \\
\text{X}_{3i} & = \text{Observational Learning}_i \\
\text{X}_{4i} & = \text{Expectations}_i \\
\text{X}_{5i} & = \text{Self-Regulation}_i \\
\text{X}_{6i} & = \text{Behavioral Capability}_i \\
\text{X}_{7i} & = \text{Self-Efficacy}_i \\
u_j & \sim \mathcal{N}(0, \sigma_u^2) : \\
    & \quad \text{random intercept for iteration } j \\
\varepsilon_{ijt} & \sim \mathcal{N}(0, \sigma_\varepsilon^2) : \\
    & \quad \text{residual error term}
\end{aligned}
\]

\subsubsection{Model 2: Temporal Development Model}\label{app:model2}

The temporal development model extends Model 1 by incorporating interaction terms:
\begin{align}
y_{ijt} =\ & \beta_0 + \beta_1\, \text{C}_{ijt} + \sum_{k=2}^{7} \beta_k\, \text{X}_{ki} \notag \\
           & + \sum_{k=8}^{13} \beta_k\, \left(\text{X}_{(k-6)i} \times t\right) + u_j + \varepsilon_{ijt}
\end{align}

\noindent where the interaction terms are defined as:
\begin{align*}
\beta_8 \left(\text{X}_{2i} \times t\right) & = \text{Reinforcements} \times \text{Round} \\
\beta_9 \left(\text{X}_{3i} \times t\right) & = \text{Observational Learning} \times \text{Round} \\
\beta_{10} \left(\text{X}_{4i} \times t\right) & = \text{Expectations} \times \text{Round} \\
\beta_{11} \left(\text{X}_{5i} \times t\right) & = \text{Self-Regulation} \times \text{Round} \\
\beta_{12} \left(\text{X}_{6i} \times t\right) & = \text{Behavioral Capability} \times \text{Round} \\
\beta_{13} \left(\text{X}_{7i} \times t\right) & = \text{Self-Efficacy} \times \text{Round}
\end{align*}

\noindent The estimated equation for Model 2 is:
\begin{align*}
y_{ijt} =\ & -0.127 + 1.426\, \text{C}_{ijt} - 1.871\, \text{X}_{2i} + 1.397\, \text{X}_{3i} \\
           & - 1.569\, \text{X}_{4i} - 1.373\, \text{X}_{5i} - 0.124\, \text{X}_{6i} \\
           & + 2.510\, \text{X}_{7i} - 0.172\, (\text{X}_{2i} \times t) \\
           & + 0.115\, (\text{X}_{3i} \times t) - 0.211\, (\text{X}_{4i} \times t) \\
           & - 0.135\, (\text{X}_{5i} \times t) - 0.036\, (\text{X}_{6i} \times t) \\
           & + 0.318\, (\text{X}_{7i} \times t) + u_j + \varepsilon_{ijt}
\end{align*}

\noindent where $u_j \sim \mathcal{N}(0, 7.36 \times 10^{-6})$ and $\varepsilon_{ijt} \sim \mathcal{N}(0, 0.035)$.

\subsubsection{Model Comparison}\label{app:model-comparision} We used a likelihood ratio test to compare the two nested models: \begin{displaymath} \Lambda = -2(\ell_1 - \ell_2) = -2(533.44 - 733.35) = 399.82 \end{displaymath} \noindent where $\ell_1$ and $\ell_2$ are the log-likelihoods of the fixed-effects and temporal development models, respectively. Under the null hypothesis that the additional parameters in the temporal development model are zero, $\Lambda$ follows a chi-squared distribution with 6 degrees of freedom, i.e., $\Lambda \sim \chi^2_{(6)}$. The result ($p < .001$) provides strong evidence that the temporal development model significantly improves model fit, supporting our hypothesis that SCT constructs' influence evolves across successive interaction rounds.

%% file: appendix/sensitivity_analaysis.tex

To validate our experimental design and address methodological considerations, we conducted three complementary analyses: bootstrap confidence intervals, round subset sensitivity testing, and leave-one-out analysis. These analyses confirm the reliability of our findings across different statistical validation approaches.

\subsection{Bootstrap Confidence Intervals}
\begin{figure}[!htbp]
    \centering
    \includegraphics[width=\columnwidth]{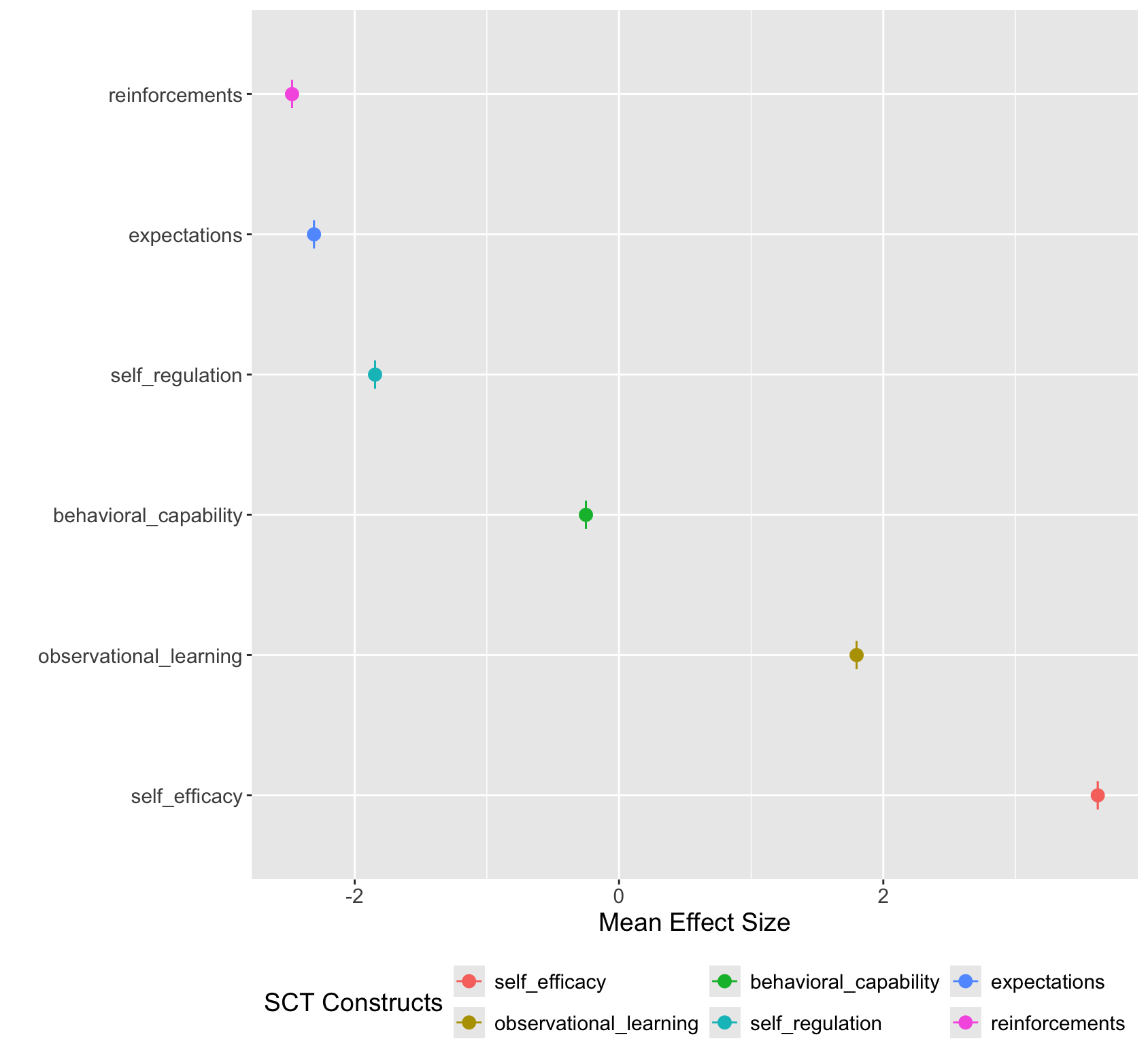}
    \caption{Bootstrap Confidence Intervals for Mean Effects}
    \label{fig:bootstrap}
\end{figure}

To validate the statistical reliability of our estimates, we conducted a bootstrap analysis with $1,000$ resamples to establish confidence intervals for the mean effects of each SCT construct (Figure~\ref{fig:bootstrap}). The results reveal remarkably narrow confidence intervals, indicating high precision in our effect size estimates despite not performing formal convergence testing. Importantly, there is clear separation between the confidence intervals of different constructs, confirming the statistical significance of the distinctions we observed. Self-efficacy shows the strongest positive effect ($3.47$, $95\%$ CI $[3.36,\ 3.58]$), followed by observational learning ($1.74$, $95\%$ CI $[1.67,\ 1.81]$). Behavioral capability remains near neutral ($0.13$, $95\%$ CI $[0.05,\ 0.21]$), while self-regulation ($-1.75$, $95\%$ CI $[-1.82,\ -1.68]$), expectations ($-2.15$, $95\%$ CI $[-2.22,\ -2.08]$), and reinforcements ($-2.38$, $95\%$ CI $[-2.46,\ -2.30]$) show progressively stronger negative effects. This bootstrap analysis confirms that our choice of $100$ repetitions provided sufficient statistical power to detect and characterize these effects reliably.

\subsection{Sensitivity of Mean Effects to Number of Rounds}

\begin{figure}[!htbp]
    \centering
    \includegraphics[width=\columnwidth]{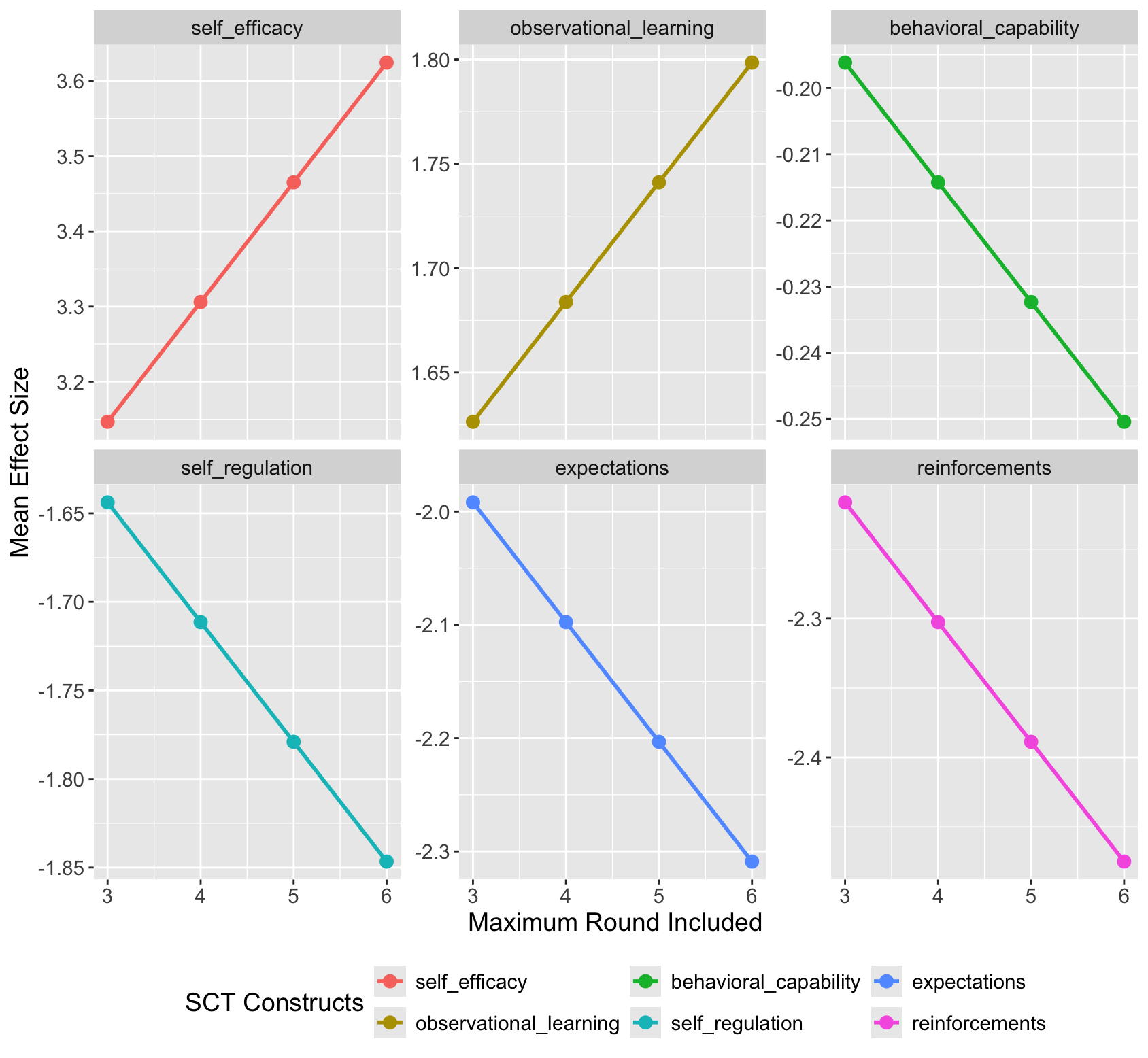}
    \caption{SCT Constructs' Sensitivity to Excluding Individual Rounds}
    \label{fig:sensitivity}
\end{figure}

To assess potential sensitivity of our results to the number of interaction rounds, we conducted a stability analysis examining how mean effect sizes change when including different numbers of rounds in our analysis (Figure~\ref{fig:sensitivity}). While the absolute magnitude of effects increases or decreases with additional rounds, the relative patterns and directional trends remain consistent across all SCT constructs. Positive constructs (self-efficacy and observational learning) consistently show positive and increasing effects regardless of how many rounds are included. Similarly, negative constructs (expectations, reinforcements, and self-regulation) maintain their negative trajectories across all subsets of rounds. Behavioral capability remains relatively neutral throughout. This stability analysis supports the robustness of our findings and suggests that our conclusions would remain consistent even with variations in the number of interaction rounds.

\subsection{Leave-One-Out Analysis}

\begin{figure}[!htbp]
    \centering
    \includegraphics[width=\columnwidth]{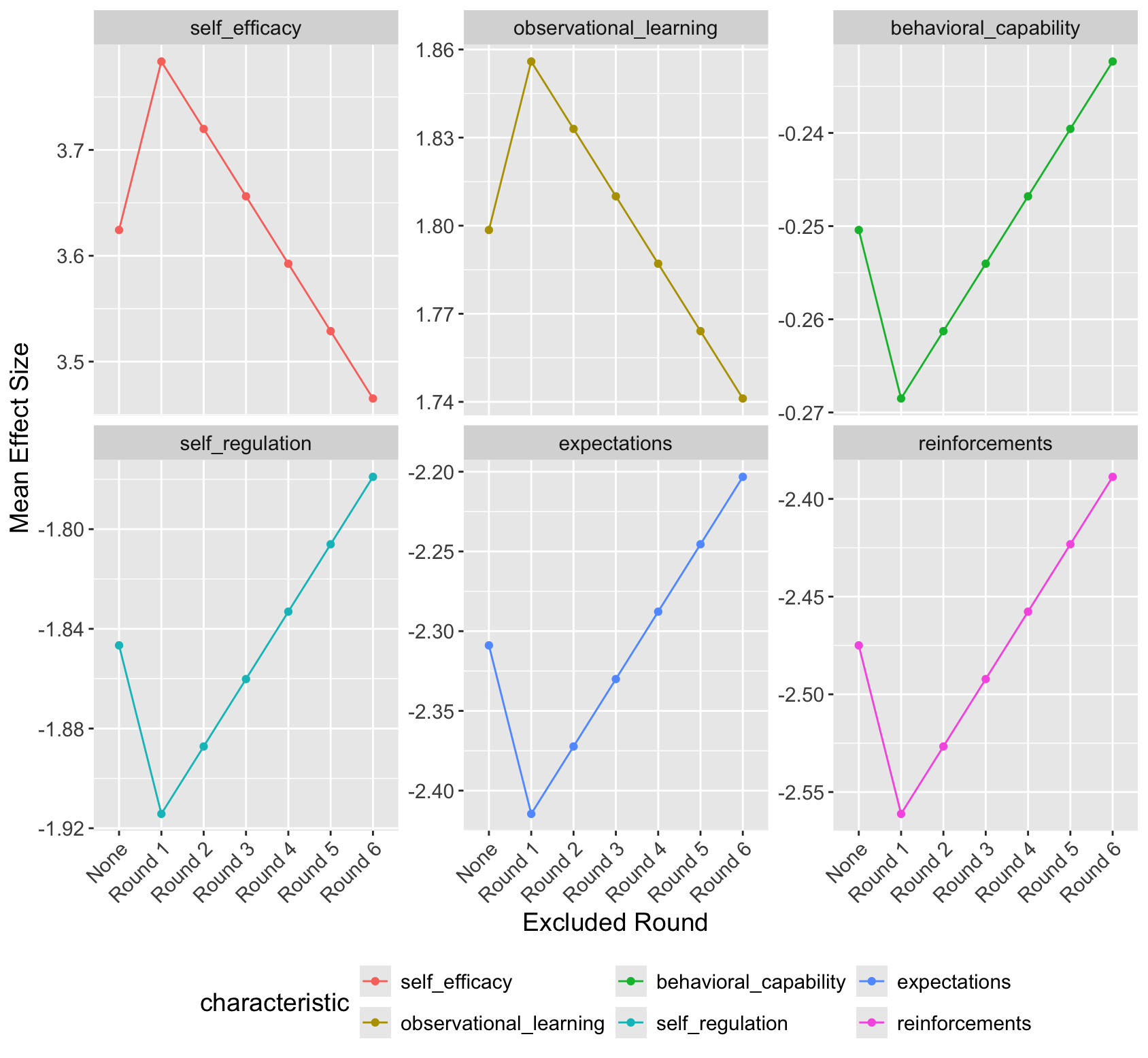}
    \caption{Sensitivity to Excluding Individual Rounds}
    \label{fig:exclusion}
\end{figure}

To examine the impact of specific interaction rounds, we conducted a leave-one-out analysis, systematically excluding each round and recalculating the mean effects (Figure~\ref{fig:exclusion}). This analysis revealed patterns showing how different rounds contributed to our findings. The magnitude of effects varied, but the relative ranking and directional trends of the SCT constructs remained consistent. Excluding Round 1 led to the most substantial deviations, suggesting it captures important baseline behavior. Excluding Round 6 moved effect sizes closer to zero, indicating the final round captures more pronounced manifestations after multiple interactions. These patterns suggest a gradual strengthening of construct manifestation over successive rounds, not random fluctuations. This analysis supports our findings’ robustness, showing they’re not disproportionately influenced by any single round. Each round contributes to a coherent progression of agent behavior consistent with our theoretical framework.